\documentclass[runningheads]{llncs}
\usepackage[T1]{fontenc}
\usepackage[]{todonotes}

\usepackage{graphicx}

\usepackage{subcaption}
\begin{document}
%
\title{(Demo) Systematic Experimentation Using Scenarios in Agent Simulation: Going Beyond Parameter Space}

%
\author{Vivek Nallur\inst{1}\orcidID{0000-0003-0447-4150} \and
Pedram Aghaei\inst{1}\orcidID{0009-0008-6316-5335} \and
Graham Finlay\inst{2}\orcidID{0000-0002-4798-2393}}
\authorrunning{V. Nallur et al.}
%
\institute{School of Computer Science, University College Dublin, Ireland \\
              \email{vivek.nallur@ucd.ie | i.pedramaghaei@gmail.com} 
\and
School of Politics and International Relations, University College Dublin, Ireland\\
              \email{graham.finlay@ucd.ie}
}
\maketitle              
\begin{abstract}
This paper demonstrates a disconnected ABM architecture that enables domain experts, and non-programmers to add qualitative insights into the ABM model without the intervention of the programmer. This role separation within the architecture allows policy-makers to systematically experiment with multiple policy interventions, different starting conditions, and visualizations to interrogate their ABM.

\end{abstract}
\section{Introduction}
The ideal that agent-based modelling (ABM) in social simulation strives to achieve, in many cases, is a true representation of the `society-of-agents' under study, so that we may gain insight into (or even generate) surprising interactions, emergent behaviour, and some level of explainability in an otherwise complex scenario. This promise has led ABM to be used in many and varied domains, e.g., GIS and socio-ecological modelling~\cite{heppenstall2011agent}\cite{filatova2013spatial}, migration networks~\cite{smith_modelling_2014}\cite{klabunde_computational_2018}, epidemiological and crisis simulation~\cite{roche_agent-based_2011}\cite{kruelen_how_2022}, computer games~\cite{meyers_computer_2012}, pedestrian dynamics ~\cite{alqurashi_multi-class_2017}\cite{karbovskii_multimodel_2018}, self-adaptive software~\cite{nallur_clonal_2016}\cite{song2015architectural}, modelling emergence\cite{nallur2016algorithm}, emotion modelling~\cite{horned_models_2023}\cite{amigoni_anxiety_2024}.

Unfortunately, agent-based modelling mechanisms are rarely built to accommodate multiple different experts. To add an additional wrinkle, the output of the model sometimes needs to be interpreted or evaluated by a completely different expert. 
The development and utilization of agent-based models (ABMs) often requires the acquisition of expertise across multiple domains. A common challenge faced by agent-modelers is that many aspects of the phenomena being modeled are typically described qualitatively rather than quantitatively. Depending on the discipline, translating these qualitative concepts into the parameters or rules of a computational model can be extremely difficult. The tension between qualitative descriptions and quantitative modeling presents an ongoing obstacle for agent-based modelers.
This paper demonstrates an architecture that allows different qualitative experts to influence and use the ABM, without the programmer intervening. Specifically, we report on the architecture of an ABM that looks at economic migrants into Ireland, and allows ethnographers to use their qualitative knowledge in shaping how the agents (migrants) behave and also allows policy experts to compare and contrast multiple policy interventions side by side, without being aware of the programmer or the ethnographer. This kind of architecture allows for a systematic exploration of complex domains, beyond typical tools such as NetLogo, Jason, Mesa, etc. In the sections that follow, we describe the architecture, and the process by which different experts can influence the model.

\section{Architecture}
The architecture is designed in order to allow different roles to participate in the ABM creation, modification and usage at different times. In the pursuit of this role separation, we propose the notion of a \texttt{BehaviourFlow} expressed visually in graphs, that enables domain-experts and non-programmers to restructure distinct aspects of the simulation model.
\subsection{Conventional Approaches}
When we simulate a real-world domain using an agent-based model, it is typically necessary to run the model multiple times to observe how the output changes in response to variations within the model. With simple models, the logic and algorithm are usually consistent, and differences are reflected in the input parameters, which are traditionally single variables that can take different values.
From a programming perspective, to compare different models, we define different sets of input variables, iterate over them, and run the model in a loop. These kinds of mechanisms are already present in current ABM tools. For example, Mesa has a module called \texttt{batch\_run} that can fulfill this need. The user can define a range or set of inputs for the model, and this module will run the model based on all combinations of these input variables.

NetLogo has a similar mechanism using \texttt{BehaviorSpace} where user can define experiments and run a model many times, systematically varying the model’s variables and recording the results of each model run. This process is sometimes called “parameter sweeping” and enables user to explore the model’s space of possible variables and determine which combinations of variables cause the result of interest. Again, the variation of models are limited to combination of input variables, and the modeller would define changes in behaviour depending on the combination of variables chosen. After running a BehaviorSpace, the output is be a spreadsheet or a table. The user is expected to use another tool to analyse and gain insights from the output.

However, for complex models, users may need to compare the effects of more complicated distinctions rather than just a single input variable. These distinctions could involve a combination of related input variables or different variations in the model's logic.
Conventionally, there are two options: either manually write and run different models separately for comparison, or write a comprehensive model and hardcode different variations as a selector input element. Even after doing this, a manual systematic approach is required to compare outputs from different combinations.
\subsection{Current Architecture Implementation}
Our proposed framework is built using Python programming language that enables us to exploit the popularity and versatility of Python and its rich package availability. The ABM implementation is built on top of the Mesa and Mesa-Geo packages\footnote{https://pypi.org/project/Mesa/}. So, all core components of these packages like the scheduler and data collector are available in this framework
Here we introduce a new concept called \texttt{Scenario} which consists of set of \texttt{Policies} and one \texttt{BehaviourFlow} for each agent type in the model, in addition to conventional input variables. A \texttt{Policy} is set of actions that will apply to each individual agent if all corresponding conditions are met. A \texttt{BehaviourFlow} is an XML file that can be visualized using any graph/network diagram editor (such as the freely available \texttt{yED}\footnote{https://www.yworks.com/products/yed/download}) and define the sequence and relation of behaviours for each agent type.
With this implementation, a Scenario becomes a single input component for model. The Scenario is a JSON file, which can be completely designed using Web-based UI. This mechanism enables policymakers to construct different Scenarios and systematically compare the complex interplay of agent-behaviour, starting conditions, as well as interventions. The Web-based UI functions as a drop-in replacement for the Mesa visualisation server and provides adequate visualisation to compare the scenarios before starting the simulation and also comparing the results, both using charts and choropleth maps.

\section{Domain Expert Designing Scenarios}
\subsection{Defining Policy}
\begin{figure}
    \centering
    \begin{subfigure}[b]{0.49\textwidth}
        \centering
        \includegraphics[width=\textwidth]{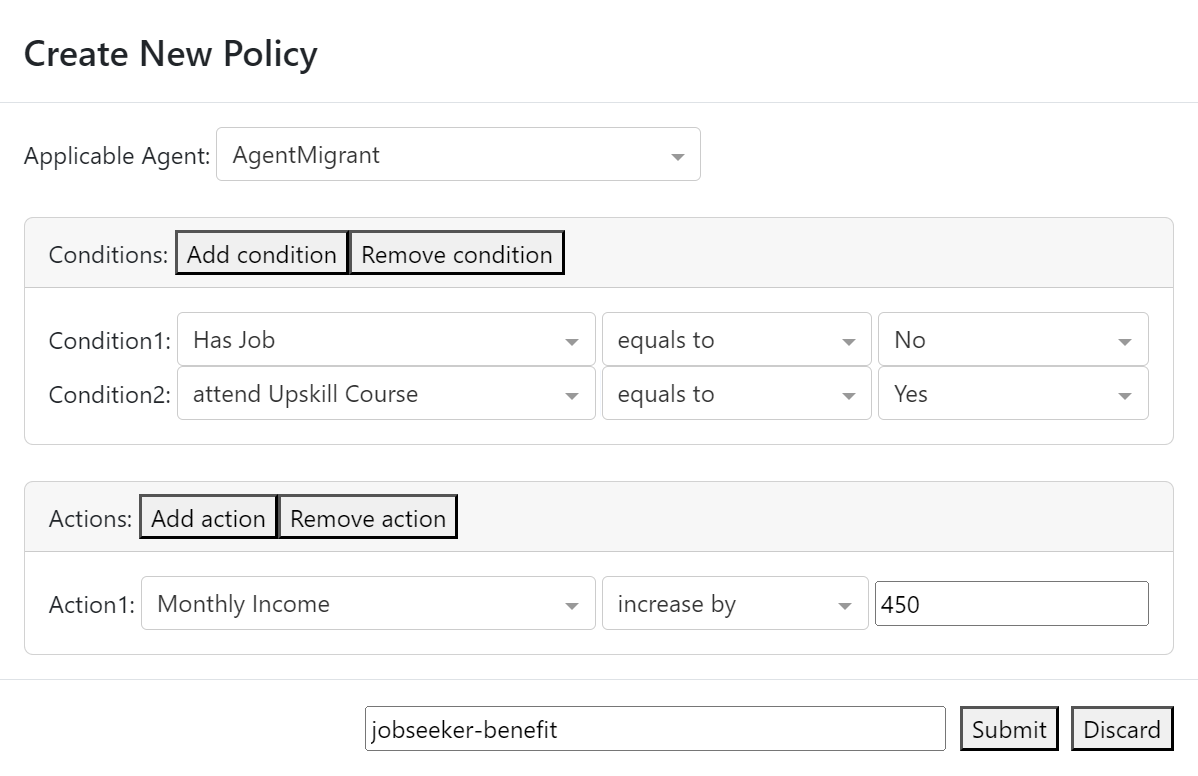}
        \caption{Jobseeker Policy}
        \label{fig:sample_policy1}
    \end{subfigure}
    \hfill
    \begin{subfigure}[b]{0.49\textwidth}
        \centering
        \includegraphics[width=\textwidth]{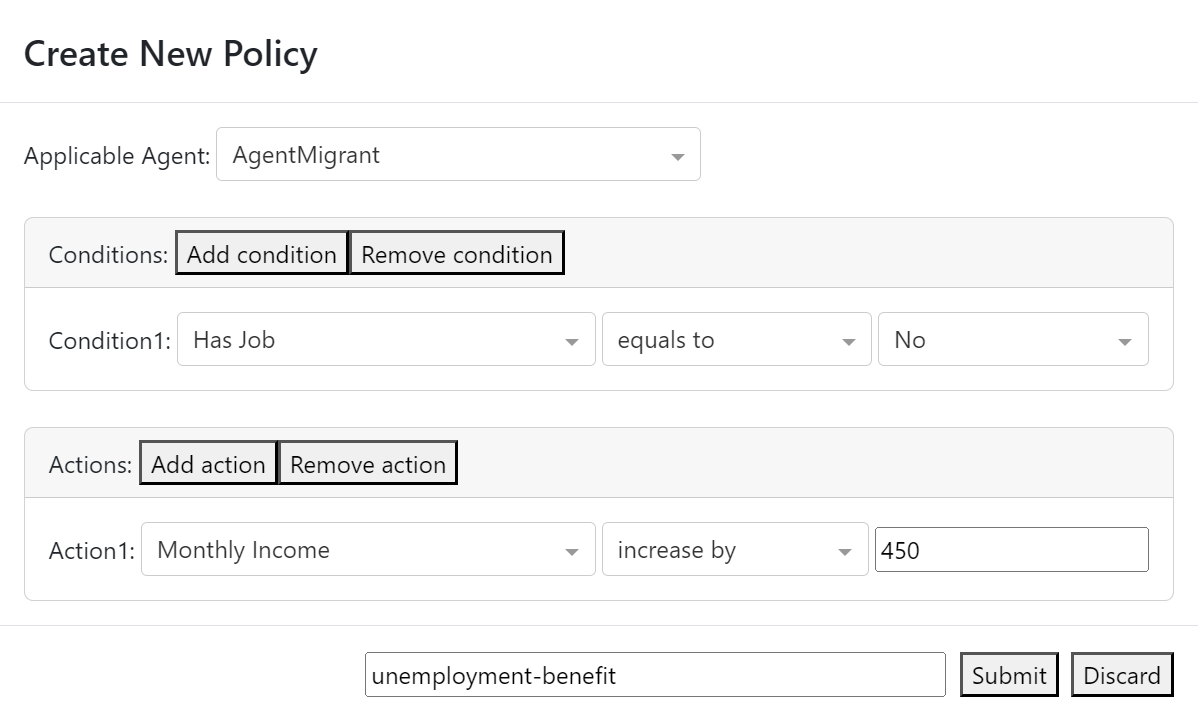}
        \caption{Unemployment Policy}
        \label{fig:sample_policy2}
    \end{subfigure}
    \caption{Two Sample Policies}
    \label{fig:sample_policies}
\end{figure}

Policies are tools that can be created by policy-makers and ABM end-users to model an intervention on a sub-population of agents. A Policy is designed to operate on units, as small as individual agents, and as large as the entire population. During each time step of the simulation, the model checks the eligibility of each policy for every agent. Once eligibility is determined, the corresponding actions are applied to all applicable agents. The policy-maker (or end-user) can create an arbitrary number of policies as the initial step in designing a scenario. Since all steps of policy definition occur within the user interface, the policy-maker can modify models with more complexity than a single parameter. Figure \ref{fig:sample_policies} shows a screenshot for defining a policy.

\subsection{Different Types of Graphs}
\begin{figure}
    \centering
    \begin{subfigure}[b]{0.40\textwidth}
        \centering
        \includegraphics[width=\textwidth]{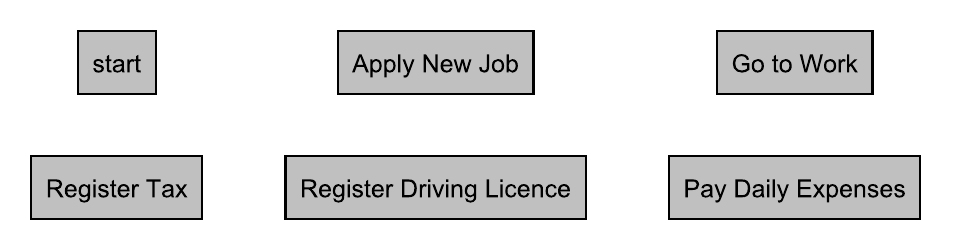}
        \caption{raw}
        \label{fig:Behaviour_raw}
    \end{subfigure}
    \hfill
    \begin{subfigure}[b]{0.10\textwidth}
        \centering
        \includegraphics[width=\textwidth]{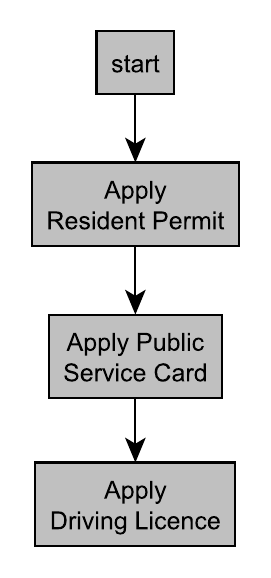}
        \caption{sequential}
        \label{fig:behaviour_sequential}
    \end{subfigure}
    \hfill
    \begin{subfigure}[b]{0.40\textwidth}
        \centering
        \includegraphics[width=\textwidth]{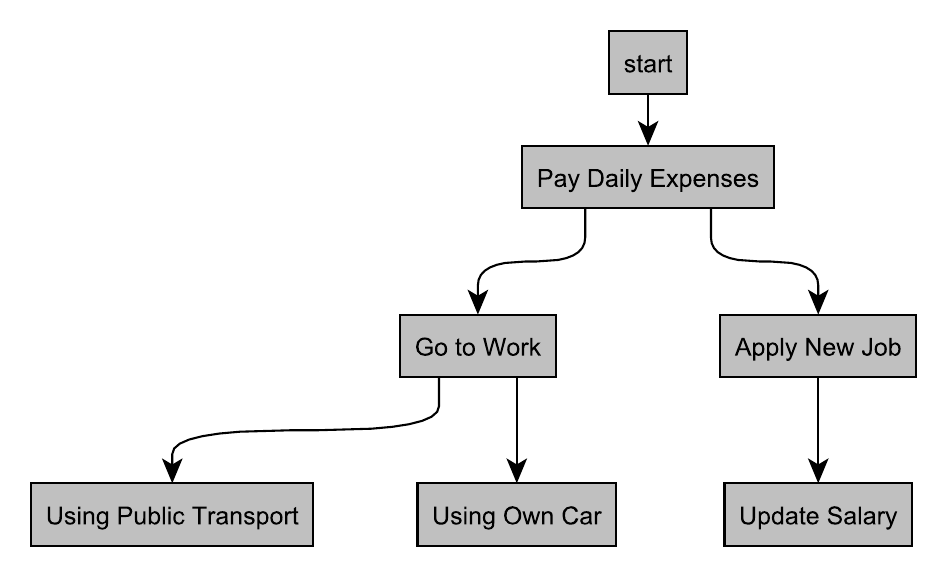}
        \caption{complex}
        \label{fig:behaviour_complex}
    \end{subfigure}
    \caption{Three Samples of BehaviourFlow}
    \label{fig:BehaviourFlow}
\end{figure}

Another extension for going beyond parameter limitation in model definition is BehaviourFlows. With this mechanism, the end user can re-arrange the behaviour of each agent type, change relations and the probability of executing each behaviour independently. This gives a remarkable customization to the end user to define new scenarios, without changing the code in runtime, rather than in the programming phase. to achieve the same result, one should hardcode every possibility in the code in applications like MEsa and NetLogo. These tools make it much easier to develop deliberative agents.
The Framework will create a raw BehaviourFlow by default (Figure \ref{fig:Behaviour_raw}) and the user can use external applications (such as yEd) to modify it~(Figure \ref{fig:behaviour_sequential} and \ref{fig:behaviour_complex})

\subsection{Running Multiple Models and Saving Results}
\begin{figure}[ht]
    \centering
    \includegraphics[width=0.25\textwidth]{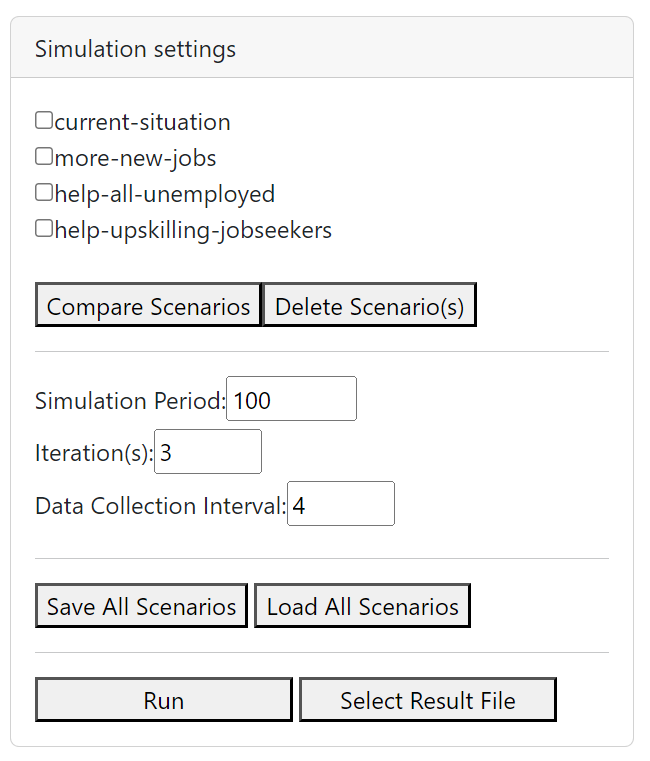}
    \caption{simulation settings}
    \vspace{-2\baselineskip}
    \label{fig:simulation_setting}
\end{figure}

\begin{figure}[ht]
    \centering
    \includegraphics[width=0.9\textwidth]{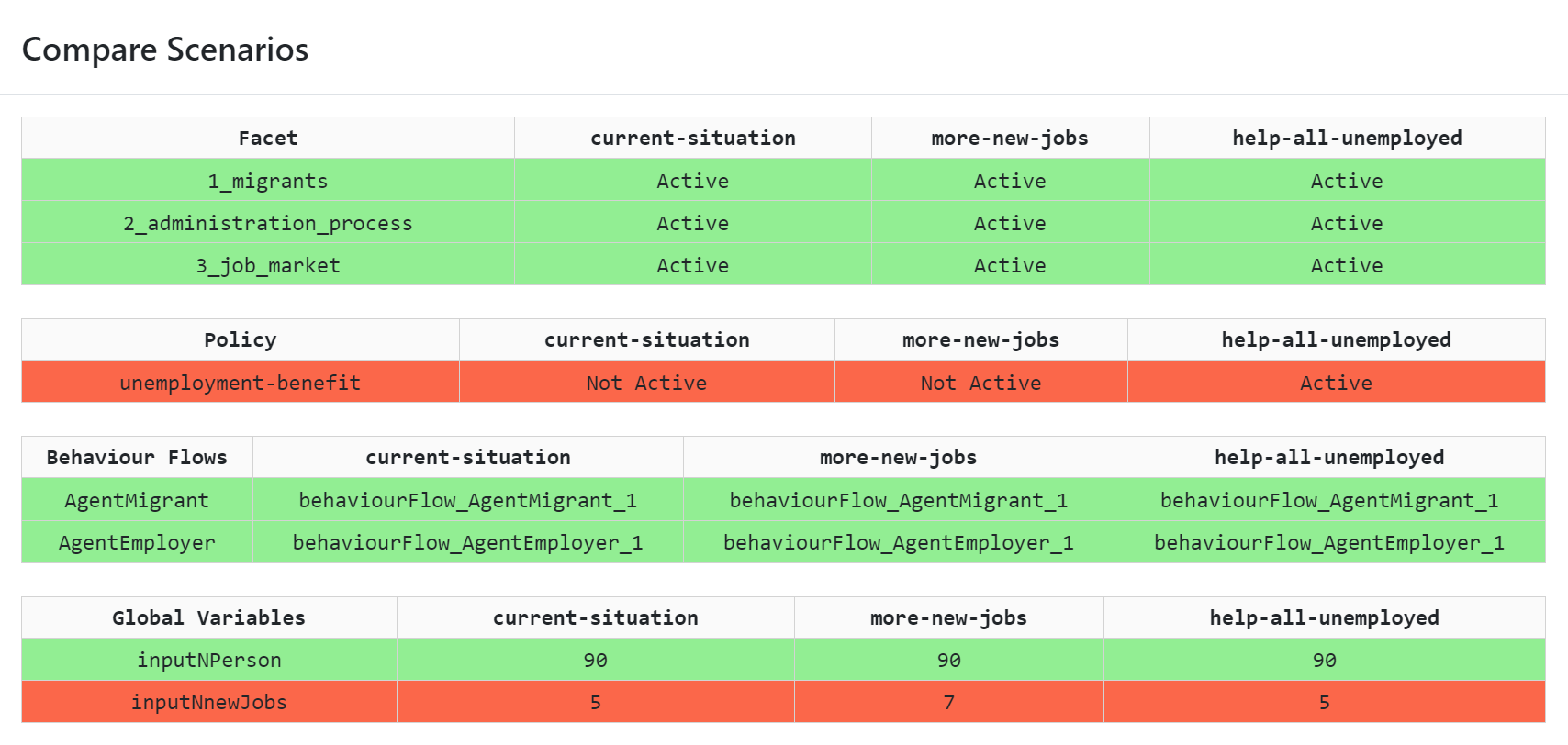}
    \caption{A table to visually compare different scenarios before starting simulation}
    \vspace{-2\baselineskip}
    \label{fig:scenario_compare}
\end{figure}

After designing the scenarios, the user can further determine the simulation settings, such as duration of the simulation, the number of iterations per scenario, and the data collection interval. These settings are common to all scenarios. Figure \ref{fig:simulation_setting} shows the simulation settings panel. As seen in the figure, a user is not only able to save-and-run multiple complex scenarios, but also save and retrieve their results for later comparison. Complex scenarios can be difficult to compare, and hence Figure~\ref{fig:scenario_compare} shows the comparison table that the user is able to consult to view
differences between scenarios.

\section{Conclusion}
In agent-based modelling where complex social scenarios are common, this disconnected architecture allows for a separation-in-time of different experts. Not only does this allow for the involvement of multiple domain experts, but it also allows qualitative insights to be integrated into the simulation, as they arrive. That is, the ABM is not frozen by the behaviours conceived of, during design-time. Due to lack of space, we have omitted the mention of details of how this is enabled, however, the code for the simulation tool built using this architecture is available, as open-source~\footnote{\url{https://csgitlab.ucd.ie/vivek/cothrom/-/tree/faceted-behaviour}}

%
%
\bibliographystyle{splncs04}
\bibliography{references}
\end{document}